\documentclass[12pt]{article}
\begin{document}
\begin{flushright}
{\tt hep-ph/0203126}\\
OSU-HEP-02-03 \\
March, 2002 \\
\end{flushright}
\vspace*{2cm}
\begin{center}
{\baselineskip 25pt
\large{\bf 
Orbifold Breaking of Left-Right Gauge Symmetry
}}

\vspace{1cm}

{\large Yukihiro Mimura\footnote
{email: {\tt mimura@okstate.edu}}
and S. Nandi\footnote
{email: {\tt shaown@okstate.edu}}
}
\vspace{.5cm}

{\small {\it Physics Department, Oklahoma State University,
             Stillwater, OK 74078}}

\vspace{.5cm}

\vspace{1.5cm}
{\bf Abstract}
\end{center}

We study the five dimensional left-right symmetric gauge theory 
in which the gauge symmetry is broken down through orbifold compactification to four dimensions.
The right handed charged gauge boson, $W_R$ is a Kaluza-Klein excited state 
in this model, and unlike the usual left-right symmetric model in four dimension,
there is no mixing in the charged sector. We develop the formalism and give expressions
for the gauge boson masses and mixings and their couplings to the fermions.
Phenomenology of the model is briefly studied to put constraints on the
right-handed gauge boson masses and the orbifold compactification scale.

\bigskip
\newpage

\section{Introduction}
\baselineskip 20pt

The Standard Model (SM) in four dimensions has been well established as an 
effective theory below the weak scale.
There have been many attempts to go beyond the SM to understand several questions
not answered in the SM framework.
One attempt is the extension of the gauge group,
while the other being the extension of the number of dimensions beyond the usual
four space-time as motivated by the string theory.

In four dimensions, minimal extension of the SM gauge symmetry is to left-right symmetry,
$SU(3)_c \times SU(2)_L \times SU(2)_R \times U(1)_{B-L}$ \cite{Pati:1974yy, Mohapatra:1979ia}.
This extension gives natural explanations \cite{Mohapatra:1983aa} 
of two important questions not explained in the SM,
the origin of parity violation at low energy weak interaction and also the non-vanishing
of the neutrino masses as the current data strongly favor.
In addition to the extended gauge symmetry, the theory also assumes discrete left-right symmetry
{\it i.e.} the Lagrangian is invariant under the exchange of the left and right handed matter multiplets.
In the four-dimensional theory, the gauge symmetry is broken by using suitable Higgs multiplets.

Recent realization that the string scale may be much smaller than the Planck scale,
even as low as few tens of TeV has inspired study of gauge symmetry in higher dimensions
with the subsequent compactification to four dimension \cite{Witten:1996mz,Arkani-Hamed:1998rs}.
Standard Model and its minimal supersymmetric extension has been formulated in five dimensions
with compactification at an inverse TeV scale and their phenomenological consequences 
have been explored \cite{Pomarol:1998sd,Barbieri:2000vh}.
In these studies, the usual Higgs mechanism has been used to break the gauge symmetry.

Compactification of this higher dimensional theory to four dimensions using a suitable 
manifold or orbifold, and choice of suitable boundary conditions,
opens up a new mechanism for breaking gauge symmetry \cite{Scherk:1978ta}, an interesting alternative to the
usual Higgs mechanism.
A symmetry breaking occurs when the different components in a multiplet of the gauge group $G$
is assigned different quantum number for the discrete symmetry group of the 
compactifying orbifold \cite{Kawamura:1999nj}.
Such orbifold breaking of $SU(5)$ and $SO(10)$ Grand Unified Theories (GUT) and their implications
as well as the symmetry breaking patterns of various group under orbifold compactifications
have been studied \cite{Kawamura:2000ev, Hebecker:2001jb}.
In these studies, orbifold compactifications have been used to break the GUT symmetry to the SM.

In this work, we study the minimal extension of the SM, the left-right gauge model,
$SU(2)_L \times SU(2)_R \times U(1)_{B-L}$ in five dimension.
The gauge symmetry is broken to $SU(2)_L \times U(1)_R \times U(1)_{B-L}$
upon compactification on a $S^1/Z_2 \times Z_2^\prime$ orbifold.
This is the minimal application of the orbifold breaking of a realistic gauge model.
The next stage of symmetry breaking to the SM, $SU(2)_L \times U(1)_Y$ is achieved by using
usual Higgs mechanism.
We assume only the gauge and the Higgs bosons propagate into the extra dimensions,
while the fermions are confined in the D3 brane at an orbifold fixed point.
We develop the formalism, and work out the gauge boson mass matrices and their mixing and their coupling
to the fermions.
There are several interesting differences with the usual four dimensional left-right model.
The most important qualitative differences are that the right-handed charged $W$ boson is a 
Kaluza-Klein (KK) excited state and there is no left-right mixing in the charged gauge boson sector.
Phenomenological constraints allow the right-handed neutral gauge boson mass 
as low as a TeV and the compactification scale as low as 2 TeV.
Thus, these new gauge bosons as well as their KK excitations can be explored at the upcoming LHC.


\section{Formalism: Orbifold breaking of Left-Right gauge symmetry}

We start with the 5-dimensional $(x^\mu,y)$ left-right gauge group
$SU(2)_L \times SU(2)_R \times U(1)_{B-L}$ with coupling $g_L$, $g_R$, $g^\prime$
and impose  the discrete left-right symmetry $L \leftrightarrow R$ so that
$g_L = g_R \equiv g$.
The fifth space-like dimensional coordinate $y$ is compactified in an orbifold
$S^1/Z_2 \times Z_2^\prime$. 
The orbifold is constructed by identifying
the coordinate with $Z_2$ transformation $y \rightarrow -y$
and $Z_2^\prime$ transformation $y^\prime \rightarrow -y^\prime$,
where $y^\prime = y + \pi R/2$.
Then the orbifold space is regarded as a interval $[0, \pi R/2]$
and 4 dimensional wall are placed at the folding point $y=0$ and $y=\pi R/2$.
We assume that the gauge bosons propagate into the extra dimensions,
while the fermions are confined to the 4-dimensional wall place at the
orbifold fixed points $y=0$ or $\pi R/2$.
The fermion representations are the same as in the usual left-right theory.
Using the orbifold compactification,
we break the gauge symmetry to $SU(2)_L \times U(1)_R \times U(1)_{B-L}$.

We impose the following transformation property for the five dimensional
gauge fields under $Z_2 \times Z_2^\prime$. 
\begin{eqnarray}
W_\mu(x^\mu, y) &\rightarrow& W_\mu (x^\mu, -y) = P W_\mu(x^\mu,y) P^{-1}, \\
W_5(x^\mu, y) &\rightarrow& W_5 (x^\mu, -y) = - P W_5(x^\mu,y) P^{-1}, \\
W_\mu(x^\mu, y^\prime) &\rightarrow& W_\mu (x^\mu, -y^\prime) = P^\prime W_\mu(x^\mu,y^\prime) P^{\prime -1}, \\
W_5(x^\mu, y^\prime) &\rightarrow& W_5 (x^\mu, -y^\prime) = - P^\prime W_5(x^\mu,y^\prime) P^{\prime -1}.
\end{eqnarray}
Note that the five dimensional Lagrangian is invariant under the above transformations.
For the $SU(2)_L$ and $U(1)_{B-L}$ gauge fields, we chose 
$P=P^\prime = I = {\rm diag}(1,1)$,
whereas for the $SU(2)_R$ gauge fields, $W$, we chose $P= {\rm diag}(1,1)$, $P^\prime = {\rm diag}(1,-1)$.
The five dimensional $SU(2)_R$ gauge filed $W$ can be written as
\begin{equation}
W = \frac1{\sqrt2}
\left(
\begin{array}{cc}
W^0 & \sqrt2 W^+ \\
\sqrt2 W^- & -W^0
\end{array}
\right).
\end{equation}
Then, we find that the gauge fields have following parities under $Z_2 \times Z_2^\prime$,
\begin{eqnarray}
W^0_\mu : (+,+) &,& W^\pm_\mu : (+,-), \\
W^0_5   : (-,-) &,& W^\pm_5 : (-,+).
\end{eqnarray}
The fields are Fourier expanded as
\begin{eqnarray}
W^0_\mu(x^\mu, y) &=& \sqrt{\frac2{\pi R}} \left( W^{0(0)}_\mu (x^\mu)
+ \sqrt2 \sum_{n=1}^{\infty} W^{0(n)}_\mu (x^\mu) \cos \frac{2ny}R \right), 
\label{fourier:W0} \\
W^\pm_\mu(x^\mu, y) &=& 
\frac2{\sqrt{\pi R}} \sum_{n=0}^{\infty} W^{\pm(n)}_\mu (x^\mu) \cos \frac{(2n+1)y}R, 
\label{fourier:W+} \\
W^\pm_5(x^\mu, y) &=& 
\frac2{\sqrt{\pi R}} \sum_{n=0}^{\infty} W^{\pm(n)}_5 (x^\mu) \sin \frac{(2n+1)y}R, \\
W^0_5(x^\mu, y) &=& 
\frac2{\sqrt{\pi R}} \sum_{n=0}^{\infty} W^{0(n)}_5 (x^\mu) \sin \frac{(2n+2)y}R.
\end{eqnarray}
Then we find that the 4 dimensional right handed gauge bosons $W^{0(n)}_\mu$, $W^{\pm(n)}_\mu$ have masses
${2n}/R$ and $(2n+1)/R$, respectively. 
The important thing is that $W^{\pm(n)}_\mu$ do not have massless mode,
hence, the $SU(2)_R$ symmetry is broken down to $U(1)_R$ symmetry in 4 dimension.
This orbifold breaking is the essential ingredient of this new formulation of the left-right
gauge model leading to several new interesting consequences.

We break the remaining symmetry to the SM, $SU(2)_L \times U(1)_Y$, using the Higgs mechanism.
The appropriate Higgs multiplets for this scenario are
$\chi_R (1,2,\frac12)$, $\chi_L (2,1,\frac12)$, $\Phi_1 (2,2,0)$ and $\Phi_2 (2,2,0)$.
Their transformation property under $Z_2 \times Z_2^\prime$ are
\begin{eqnarray}
\chi_R (x^\mu, y) &\rightarrow& \chi_R(x^\mu, -y) = P \chi_R (x^\mu,y), \\
\chi_R (x^\mu, y^\prime) &\rightarrow& \chi_R(x^\mu, -y^\prime) = -P^\prime \chi_R (x^\mu,y), \\
\Phi_1 (x^\mu, y) &\rightarrow& \Phi_1(x^\mu, -y) = \Phi_1(x^\mu, y) P^{-1}, \\
\Phi_1 (x^\mu, y^\prime) &\rightarrow& \Phi_1(x^\mu, -y^\prime) = \Phi_1(x^\mu, y^\prime) P^{\prime -1}, \\
\Phi_2 (x^\mu, y) &\rightarrow& \Phi_2(x^\mu, -y) = \Phi_2(x^\mu, y) P^{-1}, \\
\Phi_2 (x^\mu, y^\prime) &\rightarrow& \Phi_2(x^\mu, -y^\prime) = -\Phi_2(x^\mu, y^\prime) P^{\prime -1}.
\end{eqnarray}
Note that the components fields $\chi^+$ and $\chi^0$ in the $\chi_R = (\chi^+,\chi^0)^T$ doublet
has $Z_2 \times Z_2^\prime$ parity $(+,-)$ and $(+,+)$ respectively.
Thus, while $\chi^+$ acquires a KK mass of $O(1/R)$, $\chi^0$ remains a massless mode under
orbifold compactification, and thus can be assigned a vacuum expectation value (VEV), $v$,
to break the remaining gauge symmetry.
In the same way, if we denote $\Phi_1 = (\phi_{11}, \phi_{12})$ where $\phi_{1i}$'s are two $SU(2)_L$ doublets,
then $\phi_{11}$ and $\phi_{12}$ has $Z_2 \times Z_2^\prime$ property $(+,+)$ and $(+,-)$ respectively.
Thus, $\phi_{12}$ has no massless mode, and hence can not have a VEV.
Same remark applies for the first $SU(2)_L$ doublet of $\Phi_2$.
Thus, the assignments of the VEV's of the Higgs fields are as follows:
\begin{equation}
\langle \chi_R^{(0)} \rangle = \left( \begin{array}{c} v \\ 0 \end{array} \right), \quad
\langle \Phi_1^{(0)} \rangle = \left( \begin{array}{cc}
                                      k_1 & 0 \\
                                       0  & 0 
                                      \end{array} \right), \quad 
\langle \Phi_2^{(0)} \rangle = \left( \begin{array}{cc}
                                       0 & 0 \\
                                       0  & k_2 
                                      \end{array} \right).
\label{Higgs vevs}
\end{equation}
This assignment breaks the $SU(2)_L \times U(1)_R \times U(1)_{B-L}$ to the SM.

We comment about why we need two $\Phi$ fields.
In the case of the usual left-right model, 
because there are two vacuum expectation values in the same multiplet,
we need only one $\Phi$
to produce adequate mass matrices for up- and down-type quarks.
In our model, since, there is only one vacuum expectation value in
one multiplet $\Phi$,
then we can not construct the appropriate up- and down-type quark matrices 
with just one $\Phi$ type multiplet.

\section{Gauge Boson Mass Matrices and Mixings}

The gauge boson mass matrices can be calculated using Eqs.(\ref{fourier:W0},\ref{fourier:W+})
and the VEV's in Eq.(\ref{Higgs vevs})
and integrating over $y$.
For the charged sector, we obtain,
\begin{equation}
\left( \begin{array}{cc} W_L^{+(n)} & W_R^{+(n)} \end{array} \right)
M^2_{(n),\mbox{charged}}
\left( \begin{array}{c} W_L^{-(n)} \\ W_R^{-(n)} \end{array} \right), 
\end{equation}
\begin{equation}
M^2_{(n),\mbox{charged}}= 
\left(
\begin{array}{cc}
\frac{g^2}2 k^2 + \left( \frac{2n}R \right)^2 & 0 \\
0 & \frac{g^2}2 (k^2 +v^2) + \left( \frac{2n+1}R \right)^2
\end{array}
\right),
\end{equation}
where $k^2 = k_1^2 + k_2^2$.

Note a key feature of this model.
Unlike usual left-right model, there is no left-right mixing.
This is because, in the usual left-right model,
the general assignment of VEV for $\Phi (2,2,0)$ is
\begin{equation}
\langle \Phi \rangle= \left(
\begin{array}{cc}
k & 0 \\
0 & k^\prime 
\end{array}
\right),
\end{equation}
and the left-right mixing is proportional to $k k^\prime$.
As noted before, the orbifold $Z_2 \times Z_2^\prime$ symmetry requires either $k$ and $k^\prime$
to be zero.
Note also that even the lowest lying charged right handed gauge bosons
are KK states and acquire masses
from the orbifold compactification.
For the neutral gauge bosons mass matrices, we obtain
\begin{equation}
\frac 12 \sum_{n=0}^{\infty}
\left( 
\begin{array}{ccc}
W_L^{0(n)} & W_R^{0(n)} & B^{(n)}
\end{array}
\right)
M^2_{(n),\mbox{neutral}}
\left(
\begin{array}{c}
W_L^{0(n)} \\ W_R^{0(n)} \\ B^{(n)}
\end{array}
\right),
\end{equation}
\begin{equation}
M^2_{(n),\mbox{neutral}}
=
\left(
\begin{array}{ccc}
\frac{g^2}2 k^2 + \left(\frac{2n}R \right)^2 & -\frac{g^2}2 k^2 & 0 \\
-\frac{g^2}2 k^2 & \frac{g^2}2 (k^2 + v^2) + \left(\frac{2n}R \right)^2 & -\frac{g g^\prime}2 v^2 \\
0 & -\frac{g g^\prime}2 v^2 & \frac{g^\prime}2 v^2 + \left(\frac{2n}R \right)^2
\end{array}
\right).
\end{equation}

Note that the zero mode neutral gauge bosons (except the photon) acquire masses only from the Higgs VEV's,
while their KK excitations get masses from both the orbifold compactification as well as the Higgs VEV's.
Their is left-right mixing in the neutral sector for each level,
but there is no KK level mixing.

The mass spectra for the charged and neutral gauge bosons are obtained to be
\begin{eqnarray}
M^2_{W_1^{(n)}} &=& \frac{g^2}2 k^2 + \left(\frac{2n}R \right)^2 \quad (n\geq 0),\\
M^2_{W_2^{(n)}} &=& \frac{g^2}2 (k^2+v^2) + \left(\frac{2n+1}R \right)^2  \quad (n\geq 0), \\
&& \nonumber \\
M^2_{A^{(0)}} &=& 0, \\
M^2_{Z_1^{(0)}} &\simeq& \frac{M^2_{W_1^{(0)}}}{\cos^2 \theta_W} 
                \left(1 - (1-\tan^2 \theta_W)^2 \frac{k^2}{v^2} \right),\\
M^2_{A^{(n)}} \simeq M^2_{Z_1^{(n)}} &\simeq& \left(\frac{2n}R \right)^2 \quad (n \geq 1),\\
M^2_{Z_2^{(n)}} &\simeq& \left(\frac{2n}R \right)^2 + \frac{\cos^2 \theta_W}{2 \cos 2\theta_W} g^2 v^2
\quad (n\geq 0),
\end{eqnarray}
where $\theta_W$ is a weak mixing angle,
\begin{equation}
\tan \theta_W = \frac{g_Y}{g}, \quad g_Y = \frac{g g^\prime}{\sqrt{g^2 + g^{\prime 2}}}.
\end{equation}

For the neutral sector, we express the mass eigenstates in terms of the current eigenstates
\begin{equation}
\left(
\begin{array}{c}
Z_1^{(n)} \\ Z_2^{(n)} \\ A^{(n)}
\end{array}
\right)
=
V
\left(
\begin{array}{c}
W_L^{0 (n)} \\ W_R^{0 (n)} \\ B^{(n)}
\end{array}
\right).
\end{equation}
We obtain the mixing matrix,
\begin{equation}
V =
\left(
\begin{array}{ccc}
\cos \phi & -\sin \phi & 0 \\
\sin \phi & \cos \phi & 0 \\
0& 0 & 1 
\end{array}
\right)
\left(
\begin{array}{ccc}
\cos \theta_W & -\sin \theta_W \tan \theta_W & -\tan \theta_W (\cos 2\theta_W)^{1/2} \\
0 & (\cos 2\theta_W)^{1/2} \sec \theta_W & -\tan \theta_W \\
\sin\theta_W & \sin \theta_W & (\cos 2 \theta_W)^{1/2}
\end{array}
\right),
\end{equation}
where 
\begin{equation}
\sin \phi \simeq -\frac{(\cos 2\theta_W)^{3/2}}{\cos^4 \theta_W} \frac{k^2}{v^2}
\simeq -(\cos 2\theta_W)^{1/2} \frac{M^2_{Z_1^{(0)}}}{M_{Z_2^{(0)}}^2}.
\label{left-right mixing}
\end{equation}

\section{Gauge Bosons-Fermions Coupling}

In our model, 5D theory has a left-right symmetry
which is broken by the boundary conditions.
At the boundary, $y=\pi R/2$, the gauge fields
$W_{\mu}^\pm (x^\mu,y)$ vanish.
Thus, the orbifold fixed point
$y=\pi R/2$ does not respect the left-right symmetry.
When we include the fermions in the theory,
we have two choices: the fermions can propagate
into the bulk or they are localized at the 
4D wall orbifold fixed points.
For this analysis, we assume that the fermions are localized
at the 4D wall at the orbifold fixed point $y=0$.
For this choice, the interaction of the fermions
and the gauge bosons can be left-right symmetric.
We use the same multiplet assignments for the fermions as in the usual
left-right gauge symmetry.
Then writing the gauge interaction of the fermions in five dimensions,
supplemented by $\delta(y)$, 
and integrating over $y$, we obtain the four dimensional Lagrangian,
\begin{equation}
{\cal L}_4 = \frac{g}{\cos \theta_W}
 Z_{i\mu}^{(n)} ( A_i^{(n)} J^{\mu}_{V-A} + B_i^{(n)} J^{\mu}_{V+A}),
\label{neutral current}
\end{equation}
where 
\begin{equation}
J^\mu_{V \mp A} = \bar \psi \gamma^\mu \frac{1 \mp \gamma_5}2 \psi.
\end{equation}
We obtain the coefficient $A_i^{(n)}$ and $B_i^{(n)}$ as follows.
\begin{eqnarray}
A_1^{(0)} &\simeq& T^3_L \left(1- \frac{\sin^2 \theta_W}{(\cos 2\theta_W)^{1/2}} \sin\phi \right)
            - Q \sin^2\theta_W \left(1- \frac{\sin\phi}{(\cos2\theta_W)^{1/2}} \right), 
\label{a-1} \\
B_1^{(0)} &\simeq& -T^3_R \frac{\cos^2 \theta_W}{(\cos 2\theta_W)^{1/2}} \sin\phi
            - Q \sin^2\theta_W \left(1- \frac{\sin\phi}{(\cos2\theta_W)^{1/2}} \right), 
\label{b-1} \\
A_2^{(0)} &\simeq& T^3_L \frac{\sin^2 \theta_W}{(\cos 2\theta_W)^{1/2}} 
            - Q \frac{\sin^2\theta_W}{(\cos2\theta_W)^{1/2}}, 
\label{a-2} \\
B_2^{(0)} &\simeq& T^3_R \frac{\cos^2 \theta_W}{(\cos 2\theta_W)^{1/2}} 
            - Q \frac{\sin^2\theta_W}{(\cos2\theta_W)^{1/2}},
\label{b-2} 
\end{eqnarray}
and for $n \geq 1$,
\begin{eqnarray}
&&A_1^{(n)} = \sqrt2 A_1^{(0)}, \quad B_1^{(n)} = \sqrt2 B_1^{(0)}, \\
&&A_2^{(n)} = \sqrt2 A_2^{(0)}, \quad B_2^{(n)} = \sqrt2 B_2^{(0)}.
\end{eqnarray}
%

Note that in Eqs.(\ref{a-1},\ref{b-1}), we have set $\cos \phi \simeq 1$
and in Eqs.(\ref{a-2},\ref{b-2}), we have set $\cos \phi \simeq 1$,
$\sin \phi = 0$.


\section{Phenomenological Implications}

In this section, we briefly discuss the phenomenology of the model
as well as the present approximate bounds on the mass of the $W_2^{(0)}$,
$Z_2^{(0)}$ and the orbifold compactification scale, $1/R$.
Details and more complete analysis will be present elsewhere \cite{preparation}.
Detailed analysis for the usual left-right model using the LEP data
can be found in Ref.\cite{Erler:1999ub,Polak:pc}.
Our model has several important qualitative differences with the usual 
four-dimensional left-right model.
\begin{enumerate}
\item 
There will be tower of KK excitations both for the left and right-handed gauge
bosons. The lowest $W_2^{(n)}$ is a KK state with mass $1/R$,
whereas the first KK excitation of $W_1^{(0)}$ has a mass of $2/R$.
With the present bound on of only about 2 TeV for $1/R$ (to be discussed below),
few of these excitations might be produced at the LHC energy revealing the
higher dimensional nature of the theory.
\item
Unlike the usual left-right model,
there is no left-right mixing in the charged gauge boson sector.
Presence of such mixing can be measured in the angular asymmetry parameter
in the polarized muon decay or in the observation of 
$W_2 \rightarrow W_1 \gamma$ decay.
Observation of this mixing will be evidence against this orbifold
symmetry breaking.
\item
$W_2^{(0)}$ being a KK excited state, its coupling to the fermions
is larger by factor $\sqrt2$ than in the usual left-right model.
Thus, the production cross section as well as the decay width for $W_2^{(0)}$
will be twice of those in the usual left-right model.
Lightest $Z_2^{(n)}$ is not an KK excited state, so its production cross section
and decay width will be same as in usual left-right model.
This property can be used to distinguish easily between the two models
if $W_2$ and $Z_2$ are produced at the LHC.
\item
For our model, we have
\begin{equation}
{\cal R} \equiv \frac{M_{W_2^{(0)}}^2 \cos^2 \theta_W}{M_{Z_2^{(0)}} \cos 2 \theta_W}
= 1 + \frac{1/R^2}{\frac12 g^2 v^2}.
\end{equation}
Thus, for orbifold breaking, ${\cal R}$ is larger than one, while for the usual
left-right model, together with this Higgs breaking,
${\cal R}$ is equal to 1. Since orbifold breaking scale is higher than
the Higgs breaking in this model,
$W_2^{(0)}$ is expected to be heavier than $Z_2^{(0)}$ while in the usual
left-right model, $W_2$ is lighter than $Z_2$.
\item
Since there is no left-right mixing, the contribution to neutrinoless
double beta decay is expected to be less than in the usual left-right model.
\end{enumerate}

We now discuss briefly the bounds on $M_{W_2^{(0)}}$, $M_{Z_2^{(0)}}$ and the orbifold
compactification scale, $1/R$, for the model.
The model has five parameters $(g,g^\prime, k, v, 1/R)$.
We choose an equivalent set $(\alpha, M_{W_1^{(0)}}, M_{Z_1^{(0)}},
M_{W_2^{(0)}}, M_{Z_2^{(0)}})$.
For notational simplicity, from now on we denote
$W_1^{(0)}$, $Z_1^{(0)}$,
$W_2^{(0)}$, $Z_2^{(0)}$ by 
$W_1$, $Z_1$,
$W_2$, $Z_2$, respectively.

\subsection{Constraint on $Z_2$ mass}

We calculate the $\rho$ parameter in our model,
and use the measure value of the forward-backward
asymmetry 
for $e^+ e^- \rightarrow Z_{pole} \rightarrow e^+ e^-$
at LEP to set lower bound on $M_{Z_2}$.
$\rho$ parameter is defined by
\begin{equation}
\rho \equiv \frac{M_{W_1}^2}{M_{Z_1}^2 \cos^2\theta_W}.
\label{rho}
\end{equation}
For our model,
\begin{equation}
\rho = 1 + \Delta \rho_{\rm LR} + \Delta \rho_{\rm SM},
\label{delta-rho}
\end{equation}
where $\Delta \rho_{\rm LR} \simeq \cos 2\theta_W M_{Z_1}^2/M_{Z_2}^2$,
$\Delta \rho_{\rm SM} \simeq 3 G_F m_t^2/8 \sqrt2 \pi^2$.
The $\rho$ parameter can also be expressed as
\begin{equation}
\rho \equiv \frac{M_{W_1}^2}{\cos^2 \bar\theta_W M_{Z_1}^2}
= \frac{M_{W_1}^2}{M_{Z_1}^2} \frac{4}{3+\frac{g_V}{g_A}(1+\Delta_\phi)},
\label{rho parameter}
\end{equation}
where $\Delta_\phi \simeq -(1+\cos 2\theta_W) M_{Z_1}^2/M_{Z_2}^2$.
In deriving Eq.(\ref{rho parameter}), we have calculated 
the ratio of vector and axial-vector coupling
$g_V/g_A$ from Eq.(\ref{neutral current})
and then substituted $\cos^2 \theta_W$ in terms of $g_V/g_A$ in
Eq.(\ref{rho}).
We have also used Eq.(\ref{left-right mixing}).
Then equating Eqs.(\ref{delta-rho}) and (\ref{rho parameter}),
and using $\displaystyle A_e = \frac{2 g_V g_A}{g_V^2 + g_A^2}
= 0.15138 \pm 0.00216$ \cite{Groom:in},
we obtain
\begin{equation}
M_{Z_2} > 
 935 (970) \ {\rm GeV} \quad {\rm at} \ 95 \% (90 \%) \ {\rm CL,} 
\end{equation}
and 
\begin{equation}
\frac{g v}{\sqrt 2} > 
 780 (810) \ {\rm GeV} \quad {\rm at} \ 95 \% (90 \%) \ {\rm CL.} 
\end{equation}

\subsection{Constraint on $W_2$ mass and $1/R$}

Mass of the KK tower of the left-handed gauge boson depends on $1/R$.
These provides additional contributions to the muon decay and
thus lead to lower bound on $1/R$ \cite{Nath:1999fs,Marciano:1999ia}.
$K$-$\bar K$ mixing receives additional contribution from
the $W_1^{(m)}$-$W_2^{(n)}$ exchanges at the one loop.
This also leads to a more stringent bound on $1/R$ and hence on
$W_2^{(n)}$ mass.

Following the work of Ref.\cite{Marciano:1999ia}
and using the Fermi constant ($G_F^{(2)}$ in Ref.\cite{Marciano:1999ia}) 
for the $\overline{\rm MS}$ scheme,
we obtain for our model
\begin{equation}
1+ 0.082 \Delta_\phi + \frac{\pi^2}{12} M_{W_1}^2 R^2 = 0.99928 \pm 0.0015.
\end{equation}
This gives 
\begin{equation}
\frac1R > 1.20 (1.36) \ {\rm TeV} \ {\rm at} \ 95\% (90\%) \ {\rm CL}.
\end{equation}

We now consider the constraint from $K$-$\bar K$ mixing.
In left-right theory,
there are additional box diagrams with both
$W_1$ and $W_2$ exchanges. These diagrams
contain large logarithms of the form
$\ln (M_{W_2}^2/m_c^2)$, large LR hadronic matrix elements,
and new short distance QCD corrections \cite{Beall:1981ze}.
The effective Hamiltonian for our model is the same as in the usual
left-right model given in Ref.\cite{Beall:1981ze},
except $1/M_{W_2}^2$ is replaced by $1/\langle M_{W_2}^2 \rangle$
due to the exchanges of the KK tower with,
\begin{equation}
\frac1{\langle M_{W_2}^2 \rangle} = 2 \sum_{n=0}^{\infty} \frac1{M_{W_2^{(n)}}^2}
= \frac{\pi}2 
\frac{R^2}{\alpha} \tanh \left(\frac{\pi}2 \alpha\right),
\label{constraint:1overR}
\end{equation}
where
\begin{equation}
\alpha = R \sqrt{\frac{g^2}2 (k^2+v^2)}\ .
\end{equation}
The factor 2 comes from $\sqrt2$ factor in gauge couplings of the fermions
to the KK excited gauge bosons.
Using the bound for 
$\sqrt{\langle M_{W_2}^2 \rangle} > 1.6$ TeV from Ref.\cite{Beall:1981ze},
we obtain from Eq.(\ref{constraint:1overR}),
\begin{equation}
M_{W_2} > 2.5 \ {\rm TeV}.
\end{equation}
Assuming $\alpha<1$ $(1/R > gv/\sqrt2)$ as implicit in our pattern of symmetry breaking,
we get
\begin{equation}
\frac1R > 1.9 \ {\rm TeV}.
\end{equation}

\section{Conclusions}

We conclude summarizing our main results
We have considered the orbifold breaking of the minimal
extension of the SM gauge symmetry,
namely, left-right gauge model in five dimensions.
$SU(2)_R$ as well as the parity symmetry is broken to $U(1)_R$
via compactification to four dimensions.
We have developed the formalism as well as calculated the gauge bosons
mass matrices, mixings and their coupling to the fermions.
The new model has several distinguishing features from the
usual four dimensional left-right model as summarized in the 
beginning of section 5.
The major features being the existence of the KK excitations of the gauge bosons,
no left-right mixing in the charged sector,
lightest $W_2^{(n)}$ is a KK excitation while $Z_2^{(0)}$ is not.
We have discussed the phenomenology of the model briefly and set bounds
on the masses of $W_2^{(0)}$, $Z_2^{(0)}$ as well as the compactification scale, $1/R$.
These bounds are low enough for $W_2^{(0)}$, $Z_2^{(0)}$ and first few KK excitations of
$Z_2^{(n)}$ to be produced at the LHC.
The production cross sections as well as decay width for $W_2^{(0)}$
being different from the usual left-right model,
while not for $Z_2^{(0)}$,
is an interesting unique feature of the new model.
The detailed phenomenological study of this model,
as well as that the left-right model in five dimensions
without orbifold breaking of gauge symmetry will be presented elsewhere \cite{preparation}.

\section*{Acknowledgments}

We thank K.S. Babu, J. Lykken and R.N. Mohapatra for discussions.
We also acknowledge the warm hospitality and support of the 
Fermilab Theory Group during our participation in their Summer Visitor Program.
S.N. also thanks R.N. Mohapatra for warm hospitality during his visit to the
University of Maryland.
This work was supported in part by US DOE Grants \# DE-FG030-98ER-41076
and DE-FG-02-01ER-45684.

\end{document}